\documentstyle[psfig,twocolumn,prl,aps,amssymb]{revtex}
\begin{document}
\wideabs{
\draft

\date{\today} \title{Phase Diagram of Bilayer Composite Fermion States}
\author{V.W. Scarola and J.K. Jain}  
\address{Department of Physics, 104 Davey Laboratory, The Pennsylvania State 
University, University Park, Pennsylvania 16802}
%\vspace{.3in}\begin{minipage}[t]{5.5in} \rm
\maketitle

\begin{abstract}
We construct a class of composite fermion states for bilayer electron systems in a strong
transverse magnetic field, and determine 
quantitatively the phase diagram as a function of the layer separation,  
layer thickness, and electron density.  We find, in general, that there 
are several transitions,
and that the incompressible phases are separated by compressible ones.  The 
paired states of composite fermions, described by Pfaffian wave functions, are 
also considered.
\end{abstract}

}

\section{Introduction}

Multicomponent fractional quantum Hall 
effect is of relevance to situations when
there exists more than one species of electron, distinguished by, for example, their 
spin or the layer index.  There has been substantial interest in
multicomponent states of fractional quantum Hall effect (FQHE) in recent years 
both because of their experimental realizations,
and because of theoretical progress in the understanding of their structure. 
The focus in this article will be on multicomponent states in a bilayer system. 
For simplicity, we will assume that the Zeeman energy is sufficiently large that 
the electron spin is completely frozen, which is a good approximation in the limit
of large magnetic fields.

Multicomponent FQHE states were first considered in 1984, when Halperin\cite{Halperin}
generalized Laughlin's wave function for the $\nu=1/3$ FQHE \cite{Laughlin} 
to states containing more than one component of electron:
\begin{eqnarray}
&\chi&_{m',m'',m}[\{z_i\},\{w_r\}]=\prod_{j<k=1}^{N_1}(z_j-z_k)^{m'} 
\nonumber \\
&\times&\prod_{r<s=1}^{N_2}
(w_r-w_s)^{m''}\prod_{j,r}(z_j-w_r)^m
\nonumber \\
&\times&\exp[-\frac{1}{4}(\sum_{j=1}^{N_1}|z_j|^2+\sum_{r=1}^{N_2}|w_r|^2)]
\label{H}
\end{eqnarray}
Here, $z=x-iy$ and $w=x-iy$ denote the positions of the two components of electrons, 
labeled by subscripts $j,k$ and $r,s$.  The exponents $m''$ and $m'$ are odd 
integers, to ensure proper antisymmetry, and $m$ is an arbitrary integer.  
$N_1$ and $N_2$ are the numbers of electrons in the two layers.

These wave functions were first applied to mixed spin FQHE states, i.e., 
when electrons with both spins are relevant.
(Such a possibility arises in GaAs because the Zeeman splitting is much smaller than the 
characteristic Coulomb interaction energy for typical parameters.)
The function $\chi_{m',m'',m}$ is only the spatial part; 
the full wave function is given by 
\begin{equation}
A[\chi_{m',m'',m}[\{z_i\},\{w_r\}]\; u_1 u_2 ... u_{N_1} d_1 d_2 ... d_{N_2}]
\label{AH}
\end{equation}
where $A$ is the antisymmetrization operator and $u$ and $d$ are up and down 
spinors.  When the inter-electron interaction is spin-independent, 
the ground state is an eigenstate of spin. 
Only the wave functions with a well defined spin quantum number are valid 
variational wave functions.  (In principle, due to the spin-orbit coupling, 
other wave functions may also become relevant.  However, the spin-orbit coupling is 
weak in GaAs and almost invariably neglected in theoretical considerations.)  
Some examples of states that have a well defined spin quantum number, i.e., satisfy the Fock
condition are the fully polarized state $\chi_{1,1,1}$ at $\nu=1/3$ 
(it is identical to Laughlin's fully polarized wave function except
that the spin is pointing in the plane, giving zero z-component, rather than normal to the
plane) and the spin-singlet state $\chi_{3,3,2}$ at $\nu=2/5$.

Later, the multicomponent wave functions were also considered for bilayer systems.
The wave function in Eq.~\ref{AH} also applies to bilayer systems, but with 
$u$ and $d$ playing the role of {\em pseudo}spinors, 
indicating the occupation of the left and right layers, respectively.  When the layer 
separation $d$ is zero, the problem is equivalent to that of spinful electrons
in a single layer with the Zeeman energy set to zero, because the inter-electron interaction is 
pseudospin independent, given by 
\begin{equation}
V^{2D}(r)=\frac{e^2}{\epsilon r}
\end{equation}
independent of the layer index of the particles.
In this limit, again only a limited class of the above wave functions
has the correct symmetry.  However, for finite $d$,  the effective two-dimensional 
inter-electron interaction is explicitly pseudospin dependent: For electrons in the 
in the same layer, it is 
\begin{equation}
V_{\uparrow\uparrow}(r)=V_{\downarrow\downarrow}(r)=\frac{e^2}{\epsilon r}
\end{equation} 
but for electrons in different layers, it is 
\begin{equation}
V_{\uparrow\downarrow}(r)=\frac{e^2}{\epsilon\sqrt{r^2+d^2}}\;. 
\end{equation}
Here, $r$ is the projection of the distance in the plane and 
$\uparrow$ and $\downarrow$ denote the two layers.  
Thus, for finite $d$, none of the Halperin's wave functions can be ruled out for symmetry
reasons alone.  The simplest non-trivial state is the state $\chi_{3,3,1}$, which 
describes an incompressible state at $\nu=1/2$.  A FQHE at $\nu=1/2$ has indeed been  
observed\cite{Suen,Eisenstein} in a certain range
of layer separation in bilayer systems, believed to be well described 
by $\chi_{3,3,1}$\cite{He2}. 
(Note that the filling factor is defined to be the 
total filling; e.g., $\nu=1/2$ corresponds to a
filling factor of $1/4$ in each individual layer.) 
Another interesting example is 
the QHE state $\chi_{1,1,1}$ at $\nu=1$.  It is a non-trivial state 
for a bilayer system.  For a single layer, $\nu=1$ QHE occurs even for non-interacting
electrons, because a gap opens up due to Landau level quantization.  For the bilayer system,
however, the FQHE at $\nu=1$ originates due to interactions, as can be
seen by noting that each layer has $\nu=1/2$, for which there is no gap for 
independent electrons.  There is experimental evidence for this state also \cite{111}. 
(Distinguishing the bilayer $\nu=1$ from the single layer $\nu=1$ is subtle, due to the
unavoidable presence of some tunneling in 
the real experimental system, and a systematic study of the evolution of the state  
as a function of various parameters is necessary for its unambiguous identification. 
Further discussion and the relevant literature can be found in Ref.~\cite{Eisenstein2}.)
Recently, there has been a revival of interest in the $\nu=1$ bilayer state 
because of experimental observations \cite{Spielman} that have been interpreted in terms of a 
Josephson-like collective tunneling\cite{Wen}.

The physics of multicomponent FQHE states is expected to be intimately connected to 
and derived from the physics of the FQHE in single component systems.  However, 
Laughlin's wave function, which forms the basis of the states discussed above,
describes only a subset of the observed states at $\nu=n/(2pn\pm 1)$ in a single 
layer system.  The wave functions in Eq.~\ref{H} therefore also 
describe only a subset of a larger class of bilayer states.  
To take a simple example of states not described by $\chi_{m',m'',m}$, 
consider $m=0$, i.e., two uncoupled layers, and assume equal density in the two layers.  
The above wave function only applies to FQHE at $2/m'$, with $m'$ an odd integer, 
even though in reality FQHE will occur at a much larger class of fractions given by
$\nu=2n/(2pn\pm 1)$.
Further, Halperin's wave functions are not applicable to compressible states, which, as we shall
see, play an important role in the bilayer phase diagram.

The general theory of the FQHE in a single layer is formulated in terms of a particle called
the composite fermion, which is the bound state of an electron and $2p$ quantum mechanical
vortices\cite{Jain}.  The states of composite fermions are described by the wave 
function\cite{Jain}
\begin{equation}
\psi_{n/(2pn+1)}={\cal P}_{LLL} \Phi_{n} \prod_{j<k}(z_j-z_k)^{2p}
\end{equation}
which is interpreted as $\nu^*=n$ filled Landau levels of composite fermions, $\nu^*$ being the
filling factor of composite fermions.
Here $\Phi_{n}$ is the wave function of $n$ filled Landau levels of electrons, and the
Jastrow factor $\prod_{j<k}(z_j-z_k)^{2p}$ attaches $2p$ vortices to each electron to
convert it into a composite fermion \cite{footnote2}.  
The symbol ${\cal P}_{LLL}$ denotes the projection of
the product wave function into the lowest electronic Landau level.  
These wave functions 
are known to be extremely accurate representations of the actual wave functions. 
Laughlin's wave function is obtained for the special case $n=1$, and represents one filled
composite-fermion Landau level.
In the limit $\nu\rightarrow 1/2p$, the composite fermion filling factor approaches 
$n\rightarrow \infty$, i.e., the effective magnetic field experienced by the composite
fermions, $B^*=\rho \phi_0/n$,
vanishes.  They form a Fermi sea here \cite {Kalmeyer,HLR}, as confirmed 
experimentally\cite{CFFermisea}.
In this limit, $\Phi_{\infty}$ becomes the wave function of a Fermi sea at 
zero magnetic field, and $\psi_{1/2p}$ describes a compressible Fermi sea of composite
fermions.   It is a remarkable feature of the composite fermion theory that it
yields a description not only of all of the incompressible FQHE states, but also 
of the compressible liquids in the lowest Landau level.

The multicomponent generalization of the composite fermion theory for the mixed spin states
in a single layer has been considered extensively in the past\cite{Wu,Park,Shankar,Mandal}.
The generalized wave functions are
\begin{equation}
\psi_{n/(2pn+1)}^{(n_{\uparrow},n_{\downarrow})}
={\cal P}_{LLL} \Phi_{n_{\uparrow},n_{\downarrow}} 
\prod_{j<k}(z_j-z_k)^{2p}
\label{spin}
\end{equation}
where $n=n_{\uparrow}+n_{\downarrow}$ and $z_j$ label {\em all} particles.  
Here, $n_{\uparrow}$ and $n_{\downarrow}$ are the
numbers of occupied spin-up and spin-down composite-fermion Landau levels.
This wave function satisfies the Fock conditions for all choices of $n_{\uparrow}$ and
$n_{\downarrow}$, with each choice giving a FQHE state with a definite total spin, namely
$(N/2)(n_{\uparrow}-n_{\downarrow})/(n_{\uparrow}+n_{\downarrow})$.  
In the limit of vanishing Zeeman energy, the ground state has the smallest total 
spin: the total spin is $S=0$ when $n$ is an even integer, with  
$n_{\uparrow}=n_{\downarrow}=n/2$; when $n$ is an odd
integer, we have $n_{\uparrow}=(n+1)/2$ and $n_{\downarrow}=(n-1)/2$, with $S=N/2n$, $N$
being the total number of particles. 
As the Zeeman energy is increased, at certain critical values of the Zeeman energy 
$n_{\uparrow}$ increases by one unit  while $n_{\downarrow}$ decreases by one unit. 
At very large Zeeman energies, a fully polarized state is obtained.  This description is in
excellent agreement with transport and optical experiments\cite{Spinexpt}, and also with
exact diagonalization studies on small systems\cite{Wu}.  It has also been extended to finite
temperatures in a Hamiltonian formulation\cite{Shankar}.  It ought to be 
noted that the wave functions $\psi_{n/(2pn+ 1)}^{(n_{\uparrow},n_{\downarrow})}$
describe only uniform liquid states;
another structure, a charge/spin density waves of composite fermions, has been
discussed in the context of certain experimental anomalies\cite{Murthy}. 
Halperin's $\chi_{2p+1,2p+1,2p}$ state at $\nu=2/(4p+1)$ 
is obtained as a special case with $n_{\uparrow}=1$ and $n_{\downarrow}=1$; the other
composite fermion (CF) states are not expressible in the form of a Halperin wave function.

Our objective in this work is to generalize the composite fermion theory to bilayer systems 
and determine the phase diagram of bilayer composite fermion states.  
The generalization of the composite fermion theory from single to two layers 
is, to an extent, straightforward, as we show below. 
Much work on possible bilayer 
states has been done in the past\cite{Lopez},  but these studies do not tell us  
which states would actually occur in nature, and in what parameter range.
Our goal is to obtain a quantitative description.  We 
determine the phase diagram of composite fermion states 
as a function of two important parameters of the problem, the layer thickness $W$ and the 
layer spacing $d$ (defined to be the distance between the centers of the two 
layers).  Due to its variational nature, the study cannot rule out other 
states, 
but we believe that the states considered here exhaust the physically relevant 
uniform liquid states in the parameter ranges considered.  An important feature of the resultant phase diagram is that the incompressible CF phases are
often separated by compressible phases; i.e., as a function of the inter-layer spacing,
the transition typically occurs from a compressible state into an
incompressible state, or {\em vice versa}.  
In general, several phase transitions are possible at a given fraction. 
Many of these phase transitions occur in presently accessible experimental regimes, and
ought to be observable.

There has also been exact diagonalization work on two layer systems 
\cite{Chakraborty,Yoshioka,He,He2}.  It is restricted to extremely small systems, because of
the drastically enlarged Hilbert space (compared to
the single layer problem), and is therefore limited in its applicability.
The exact diagonalization approach is also suitable only for
incompressible states.  Nonetheless, these studies provide indications regarding the 
range of parameters where some of the states, for example $\nu=1/2$ bilayer FQHE state, are
strongest.  Our results below are generally consistent with the earlier results. 
We also compare our results with available experiments at $\nu=1/2$ and $\nu=1$.

The plan of the paper is as follows.  
In the next section, we describe the generalization of
the composite fermion theory to two layer system.
Section III  discusses the details of our computational method.  
In section IV,  we display the quantitative phase diagram of bilayer CF states 
in the density-layer separation ($\rho-d$) plane for several quantum well thicknesses
at total fillings $\nu=1/2, 1/3, 1,$ and $2/5$.  
In Section V our results are compared with exact diagonalization studies and
experiment.  Section VI considers bilayer CF states with Pfaffian structure in 
each layer.  The paper is concluded in Section VII.

\section{Composite fermion theory for bilayer systems}

In the case of a bilayer system, we already know the answer in two limits.  
When the layers are far
apart, the bilayer state is given simply by two independent single-layer 
composite fermion states.  In the other limit, when the layer
separation $d\rightarrow 0$, 
the bilayer problem is identical to the problem of spinful electrons in
a single layer (i.e., pseudospin acts like the real spin), with the 
Zeeman energy set equal to zero.  Fortunately, 
as discussed above, the nature of the CF states
is well understood in this limit.  At even numerator fractions, the ground state is a 
(pseudo)spin singlet, and at odd numerator fractions it is partially polarized.
For the pseudospin singlet state the densities in the two layers are equal; when 
the polarization is not zero, the densities depend on the direction of the total spin, 
with equal densities obtained when the total pseudospin is pointing parallel to the plane.

To investigate the intermediate situation, we consider the following class of wave 
functions which we expect to be particularly favorable energetically:
\begin{equation}
\Psi_{(\overline{\nu}_1,\overline{\nu}_2|m)}= \prod_{r,j}(z_j-w_r)^m 
\psi_{\overline{\nu}_1}[\{z_k\}] \psi_{\overline{\nu}_2}[\{w_s\}]
\label{Psi}
\end{equation}
where the fully antisymmetric wave function  
$\psi_{\overline{\nu}}$ describes a lowest Landau level projected single layer state 
at filling factor $\overline{\nu}$.
The state described by $\Psi_{(\overline{\nu}_1,\overline{\nu}_2|m)}$ will 
be denoted by $(\overline{\nu}_1,\overline{\nu}_2|m)$.  
When both $\psi_{\overline{\nu}_1}$ and $\psi_{\overline{\nu}_2}$ 
are incompressible, $(\overline{\nu}_1,\overline{\nu}_2|m)$ 
is also expected to describe an
incompressible state.  However, if one or both of $\psi_{\overline{\nu}_1}$ and
$\psi_{\overline{\nu}_2}$ are compressible,
so is $(\overline{\nu}_1,\overline{\nu}_2|m)$.  Halperin's wave functions are
obtained as special cases for $\overline{\nu}_1=1/m'$ and $\overline{\nu}_2=1/m''$.  

The physical interpretation of $(\overline{\nu}_1,\overline{\nu}_2|m)$ is as follows.
In a single layer system, the CF state is obtained by taking an electron state $\Phi_n$ and
then multiplying it by a Jastrow factor that attaches $2p$ quantum vortices to each
electron.  In the present case, we start with an electron state $\Phi_{\overline{n}_1}[\{z_j\}] 
\Phi_{\overline{n}_2}[\{w_r\}]$, i.e., a state that has 
$\overline{n}_1$ filled Landau levels in layer 1 and $\overline{n}_2$
in layer 2.  Now, because of the layer index, we have more flexibility in how vortices are
attached.  First of all, because of antisymmetry within each layer, we must attach an even
number of vortices [$2p_1$ and $2p_2$, giving 
$\overline{\nu}_1=\overline{n}_1/(2p_1\overline{n}_1+1)$ and 
$\overline{\nu}_2=\overline{n}_2/(2p_2\overline{n}_2+1)$] in the relative coordinates of a pair of electrons
within each layer.  However, the number of vortices in the relative coordinates of 
an {\em inter}-layer pair of electrons can be a third integer, which does not 
have to be even. (The pseudospin part of the full wave function 
in Eq.~\ref{AH} takes care of the 
antisymmetry with respect to the exchange of electrons in different layers.)
In other words, we can attach to each electron two types of vortices, one seen by 
other electrons in the same layer and the other type seen by electrons in the other layer.

In the symmetric gauge, the wave function describes a uniform state of electrons in two  
disks, one in each layer.  The sizes of the disks are determined by the largest powers of $z_j$
and $w_r$, which also give the number of single
particle orbitals in the lowest Landau level in the interior of the disk (apart from
an unimportant correction on the order of unity).
The largest power of $z_j$ and $w_r$ are
 $N_1/\overline{\nu}_1+mN_2$  and $N_2/\overline{\nu}_2+mN_1$, respectively.  
Here, $N_1$ and $N_2$ are
the numbers of electrons in the first and second layers, respectively.
In order to ensure that the electrons in the two layers
occupy the same area, $N_1$ and $N_2$ must be related by 
\begin{equation}
N_1 \overline{\nu}_1^{-1}+m N_2 = N_2 \overline{\nu}_2^{-1} + m N_1\;.    
\end{equation}
The filling factors of the individual layers are given by 
\begin{equation}
\nu_1^{-1}=\overline{\nu}_1^{-1}
+m \frac{N_2}{N_1}
\end{equation}
\begin{equation}
\nu_2^{-1}=\overline{\nu}_2^{-1} +m \frac{N_1}{N_2},
\end{equation}
 and the total filling factor
is $\nu=\nu_1+\nu_2$.

We specialize in the rest of the article 
to the situation when the individual densities in the two layers are 
equal, i.e., $N_1=N_2$.  From the preceding considerations, this implies 
$\overline{\nu}_1= \overline{\nu}_2\equiv \overline{\nu}$, $\nu_1=\nu_2=\nu/2$, 
and the total filling factor is given by
\begin{equation}
\nu=\nu_1+\nu_2=\frac{2\overline{\nu}}{1+m\overline{\nu}}
\end{equation} 
Since we are interested in enumerating the states at a given total filling factor $\nu$, 
we write
\begin{equation}
\overline{\nu}=\frac{\nu}{2-m\nu}
\end{equation}
Thus, for a given $\nu$, the possible states are
\begin{equation}
\left(\frac{\nu}{2-m\nu},\frac{\nu}{2-m\nu}\;|\;m\right)
\label{states}
\end{equation}

\subsection{Primary bilayer fractions}

For a single layer, the primary FQHE states $\psi_{\overline{\nu}}$ occur at fractions  
\begin{equation}
\overline{\nu}=\frac{\overline{n}}{2p\overline{n}+ 1}
\label{primary}
\end{equation}
We note here that negative values of $\overline{n}$ are also allowed \cite{footnote2}.
These produce the primary bilayer states 
\begin{equation}
\left(\frac{\overline{n}}{2p\overline{n}+ 1},\frac{\overline{n}}{2p\overline{n}+ 1}\;|\;m\right)
\end{equation}
at filling factors 
\begin{equation}
\nu=\frac{2\overline{n}}{(2p+m)\overline{n}+1}
\end{equation}
The numerator may be either even (for $|\overline{n}|\neq 1$), or one (which is possible when
$|\overline{n}|=1$).  We will only consider the primary bilayer states below.

States for which the numerator of $\nu$ is odd and greater 
than one make use of single layer states at fillings
$\overline{\nu}$ which are not primary FQHE states.  Consider for example $\nu=3/7$, for 
which the possible bilayer states are:  
\begin{equation}
\left(\frac{3}{14},\frac{3}{14}\;|\;0\right), \left(\frac{3}{11},\frac{3}{11}\;|\;1\right), 
\left(\frac{3}{8},\frac{3}{8}\;|\;2\right)
\end{equation}
These states are not likely to occur because the constituent, single layer, 
states, do not belong to a primary sequence.

Not all primary bilayer states are physical, though.  In fact, some of them can be 
ruled out by noting that the inter-layer correlations cannot be stronger than the intra-layer
correlations for any realistic situation.  Which states are ruled out is determined by
considering the $d\rightarrow 0$ limit, where the state $(\overline{\nu},\overline{\nu}|m)$ 
\begin{eqnarray}
\prod_{j,r}(z_j-w_r)^m 
\prod_{j<k}(z_j-z_k)^{2p}\psi_{\overline{n}}[\{z_j\}]  
\nonumber \\
\times \prod_{r<s}(w_r-w_s)^{2p}\psi_{\overline{n}}[\{w_r\}]
\end{eqnarray}
must be an eigenstate of the pseudospin.
Here, we have neglected the lowest Landau level projection, which is not relevant to the 
present discussion because it preserves the spin
quantum number.  As discussed earlier, this state is an eigenstate of the pseudospin 
for $m=m_c$, where 
\begin{eqnarray}
m_c=2p\;,\;\;\; \overline{n}\neq 1 \nonumber \\
m_c=2p+1\;,\;\;\; \overline{n}= 1
\end{eqnarray}
For $\overline{n}\neq 1$, the state at $m=m_c$ is pseudospin unpolarized \cite{Wu,Park}, 
whereas for $\overline{n}= 1$, the state at $m=m_c$, which corresponds to $\nu=1/(2p+1)$
is pseudospin polarized.  The latter is just the Halperin  
$\chi_{m, m, m}$ states which we know to be pseudo-spin eigenstates\cite{MacDonald}.  
The states with $m>m_c$ are clearly unphysical, because they correspond to stronger
inter-layer correlations than intra-layer correlations.

Note that a special case is the compressible state at $\nu=2/(2p+m)$, corresponding to the
limit $\overline{n}=\infty$.  Here, for $m=m_c=2p$, we have $\nu=1/2p$, where the pseudospin 
unpolarized composite-fermion Fermi sea is obtained \cite{Wu,Park}.  
(Here, the limit $n\rightarrow \infty$ can be taken along even integer values of $n$.)

The overall picture that we expect on the basis of the above considerations is as follows.
The value $m=0$ describes far separated, independent layers.  The integer $m$ 
is expected to increase in unit steps as the layer spacing is reduced, until $m=m_c$ is
reached.  This gives us the limit when the two layers are very close.
The values of $m> m_c$ are not physically relevant.

\subsection{Spherical geometry}

The spherical geometry \cite{WuYang,Haldane}
will be used in all our calculations, which is suited for an
investigation of the bulk properties of the various CF states, due to the lack of edges.
It considers $N$ electrons on the surface of a sphere of radius $R$, in the presence of a
radial magnetic field produced by a magnetic monopole at the center.  A monopole of
strength $q$, an integer or a half integer, produces a flux of $2q \phi_0$ ($\phi_0=hc/e$).
The single particle eigenstates are the monopole harmonics, $Y_{q,n,m}$
\cite{WuYang}, given by (with the binomial coefficient
${{\gamma \choose \beta}}$ is to be set equal to
zero if either $\beta >\gamma$ or $\beta<0$)
\begin{eqnarray}
&Y&_{q,n,m}(\Omega)=N_{qnm} (-1)^{q+n-m} e^{iq\phi_j} u_j^{q+m} v_j^{q-m}
\\
&\times&\sum_{s=0}^{n}(-1)^s {{n \choose s}} {{ 2q+n \choose q+n-m-s}}
 (v^*v)^{n-s}(u^*u)^s\;\;,
\label{mh}
\end{eqnarray}
where $\Omega$ represents the angular coordinates
$\theta$ and $\phi$ of the electron, and
\begin{equation}
u\equiv \cos(\theta/2)\exp(-i\phi/2)
\end{equation}
\begin{equation}
v\equiv \sin(\theta/2)\exp(i\phi/2)\;.
\end{equation}
Here, $n$ is the Landau level (LL) index ($n=0,1,...$), the orbital angular momentum 
is $|q|+n$, and $m$ is
the z-component of the angular momentum (not to be confused with the exponent in Halperin's
wave function).  The degeneracy of the lowest LL is $2|q|+1$, and it increases by two in each
successive higher LL.

The composite fermion wave function can be generalized straightforwardly to the spherical
geometry.  The $n$ filled LL state $\Phi_n$ occurs at $q_n=(N-n^2)/2n$.  The Jastrow factor
$\prod_{j,r}(z_j-w_r)$ is replaced by $\prod_{j,r}(u_j v_r-v_j u_r)$, which corresponds to
$q_J=N/2$.  Noting that the $q$ of the product is equal to the sum of the $q$'s of individual
factors,  the state 
\begin{equation}
\left( \frac{n}{2pn+ 1}, \frac{n}{2pn+1} \; | \; m\right)
\label{blstate}
\end{equation}
occurs at $Q=m q_J+2p q_1+q_n$, i.e., at  
\begin{equation}
Q=\frac{(2pn+mn+1)N-(2pn+n^2)}{2n}.
\label{Q}
\end{equation}
Indeed, the ratio of the total number of particles to the flux, $2N/2Q$, gives the 
correct filling factor $\nu=2n/(2pn+mn+1)$ in the thermodynamic limit $2N\rightarrow \infty$.

\subsection{Charge of Elementary Excitations}

We will only consider the ground state wave functions in this article, but it is
straightforward to calculate the charge of the elementary excitation of the incompressible
bilayer states.  Begin with the state in Eq.~\ref{blstate}
and add one electron to each layer, while holding the flux at the value given in Eq.~\ref{Q}.
Now the wave function is given by 
\begin{equation}
\prod_{j,r=1}^{N+1}(u_j v_r-v_j u_r)[{\cal P}_{LLL} \Phi_1^{2p}\Phi_{q'}]_L[{\cal
P}_{LLL}\Phi_1^{2p}\Phi_{q'}]_R
\end{equation}
where  $L$ and $R$ denote the left and the right layers.
The $q$'s of the inter and intra-layer Jastrow factors now increase, so the $q$ for $\Phi$,
called $q'$, must decrease to 
\begin{equation}
q'=\frac{N-n^2}{2n}-\frac{m}{2}-p
\end{equation}
in order to leave the total $Q$ invariant.  At $q'$, the total number of states in the lowest
$n$ LLs is $2q'n+n^2=N-mn-2np$, which implies that, for $N+1$ electrons, there are $2np+mn+1$
electrons in the $(n+1)^{st}$ LL.  Each one corresponds to an excited
composite fermion in the full wave
function.  Since a unit charge was added in each layer, the charge corresponding to each
composite fermion is 
\begin{equation}
e^*=\frac{e}{2np+mn+1}
\end{equation}

\section{Computational details}

Despite their simple interpretation, the wave functions of composite fermions 
are too complicated for 
exact analytical treatment.  Even for the simplest case, namely the Laughlin
wave function, it has not been possible to obtain any numbers by any analytical method.   
This is hardly surprising, because the composite-fermion wave functions describe a strongly
correlated state of electrons. 
(We note that the Chern-Simons field theoretical technique for dealing with composite
fermions \cite{Lopez2,HLR} is not able to obtain, from first principles, 
quantities like the ground state energies or the 
energy gaps.  For an approximate calculation of such quantities in a
Hamiltonian approach, we refer the reader to the literature \cite{GM}.)

Fortunately, it {\em is} possible to obtain exact results for the variational wave functions, 
using a numerical Monte Carlo method\cite{JK}. 
Quantitative numbers are obtained without making any approximations,
and the results can, in principle, 
be made as accurate as we wish, by running the Monte Carlo sufficiently long.  In practice,
reasonably accurate results, correct to 3-4 significant figures,
can be obtained for up to a total of $2N\sim$ 70 electrons.  This turns out
to be sufficient to extrapolate to the thermodynamic limit in many cases of interest 
to obtain results that can be compared directly to experiment.

In a purely two-dimensional system, the interaction is given by 
$V^{2D}(r_{jk})=\frac{e^2}{\epsilon|r_j-r_k|},$ where $\epsilon$ is the dielectric
constant of the background material.
For this interaction, the energies scale with $e^2/\epsilon l_0$, where $l_0=\sqrt{\hbar c/eB}$
is the magnetic length; the only density dependence is through the magnetic length.  
However, realistically, the electron wave function has a 
finite width normal to the plane of confinement, which
alters the form of the effective two-dimensional interaction at short
distances.  We incorporate in our calculations the effects of finite thickness by working with
an effective two-dimensional interaction given by 
\begin{equation}
V(r)=\frac{e^2}{\epsilon}\int dz_1 \int dz_2 \frac{|\xi(z_1)|^2
|\xi(z_2)|^2}{[r^2+(z_1-z_2)^2]^{1/2}}
\end{equation}
where $\xi(z)$ is the transverse wave function.  (Here, the real quantity 
$z$ is the distance along the normal direction, not to be confused with the complex
number $z_j$ denoting the position of a particle within the plane.) 
The functional form of $\xi$ is determined by
self-consistently solving the Schr\"odinger and Poisson equations, taking
into account the interaction effects through the local density
approximation including the exchange correlation potential
\cite {Meskini}.  The only input parameters are the electron density, $\rho$, and the shape of
the confinement potential, which will be assumed to be a 
square quantum well for each layer in this article.  The barrier heights of 
the wells are taken to be 276 meV in the calculation, although the results are not
significantly different from an infinite barrier.
The effect of inter-layer interaction is neglected in the LDA evaluation of $\xi$. 
For further details, we refer the reader to Ref.~\cite{Meskini}.
The self-consistent local-density approximation is expected to be accurate on the level of
20\%.  The treatment of the uniform positive background charge is somewhat complicated  
due to the finite thickness\cite{Park};  fortunately, since we are only interested in 
energy differences, we do not need to consider the neutralizing background charge explicitly.

\begin{figure*}[t]
\centerline{\psfig{figure=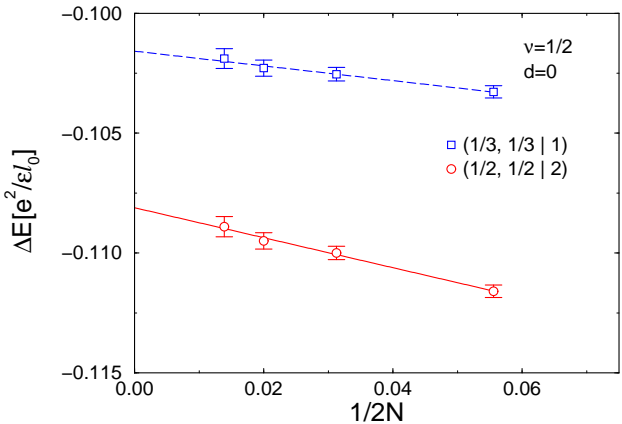,width=4in,angle=-90}}
\caption{The energies of the bilayer states $(1/3,1/3|1)$ and $(1/2,1/2|2)$ 
at $\nu=1/2$ as a function of $1/2N$, where $2N$ is the total number of
particles.  The energies are measured relative to the state $(1/4,1/4|0)$, with 
$\Delta E \equiv E_{(\overline{\nu},\overline{\nu}|m)} - 
E_{(1/4,1/4|0)}$.  The layer thickness and the layer separation are taken to be zero
in this plot.  In this and in subsequent figures the energies are given in 
units of $e^2/\epsilon l_0$, where $l_0=\sqrt{\hbar c/eB}$ 
is the magnetic length and $\epsilon$ is the dielectric 
constant of the background material.} 
\label{therm}
\end{figure*}

The filling factor is defined to be the ratio of the total number of electrons,
$2N$, to the total flux through the surface of the sphere measured in units of $\phi_0$, 
$N_\phi=2Q$.  However, for a given wave function, the ratio is $2N$ dependent, and has 
order $1/N$ deviations from its thermodynamic value, $\nu$. 
A difficulty arises because the different states that correspond to the same 
thermodynamic $\nu$  do not have the same $2N/N_\phi$ for finite systems.  
In other words, they have slightly
different densities in finite systems.  We correct for this by expressing the energies in units
of $e^2/r_0$, where $r_0$ is the interparticle separation, and then replacing $e^2/r_0$ by its
thermodynamic value.   This is equivalent to multiplying 
the distance $r$ in the expression of the interelectron interaction by the factor
$\sqrt{\rho_N/\rho}$, where $\rho_N$ is the density of the finite system.
We find that, after correcting for the density in this manner, 
the energy differences are only weakly dependent on $N$ and give reliable thermodynamic
extrapolations, as shown in Fig.~(\ref{therm}).
(In fact, the density correction makes only a small correction to the energy differences for 
the large systems considered.) The distance between the particles is measured along the 
chord.

The energy of the composite-fermion wave functions will be evaluated by Monte Carlo method,
with the lowest LL projection handled by the standard
method described in the literature \cite{JK}.  We compute the energy per particle 
to within $~0.01\%$ using $10^7$ Monte Carlo iterations.  This is done 
for several system sizes up to $2N=72$.

\section{Results and discussion}

We have evaluated the energies of various candidate ground 
states at $\nu=$ 1/3, 2/5, 1/2, and 1 as a function
of $N$, and determined the thermodynamic value of the energy difference.  The energy difference
has been calculated relative to the uncorrelated state $\Psi_{(\nu/2,\nu/2|0)}$ in each case. 
We have studied a large class of confinement potentials, but will present below 
results only for three quantum well thicknesses as a function of the electron density.  The distance $d$ separating the layers is measured from center to 
center;  its minimum value is therefore equal to the width of each quantum 
well layer.

\begin{figure*}[t]
\centerline{\psfig{figure=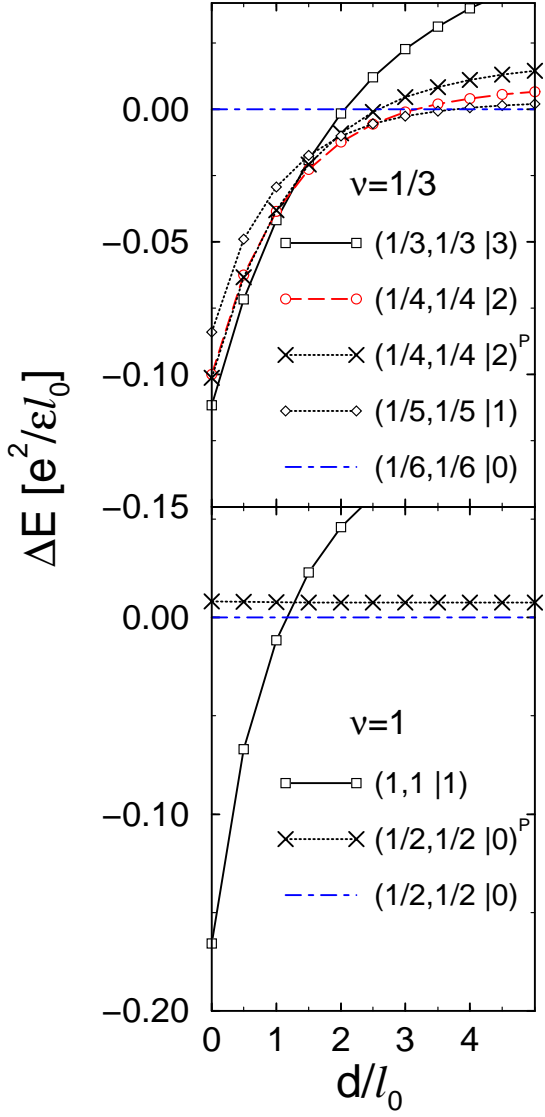,width=5in,height=7in,angle=0}}
\caption{The thermodynamic energies of various bilayer composite-fermion states at $\nu=1/3$ 
and 1, measured relative to the
energy of the uncorrelated state $(\nu/2,\nu/2|0)$ as a function of layer separation.
Each layer is taken to be strictly two-dimensional 
in this plot.  The various states $(\overline{\nu},\overline{\nu}|m)$ 
are explained in the text.}  
\label{coul_1_3_1}
\end{figure*}

\begin{figure*}[t]
\centerline{\psfig{figure=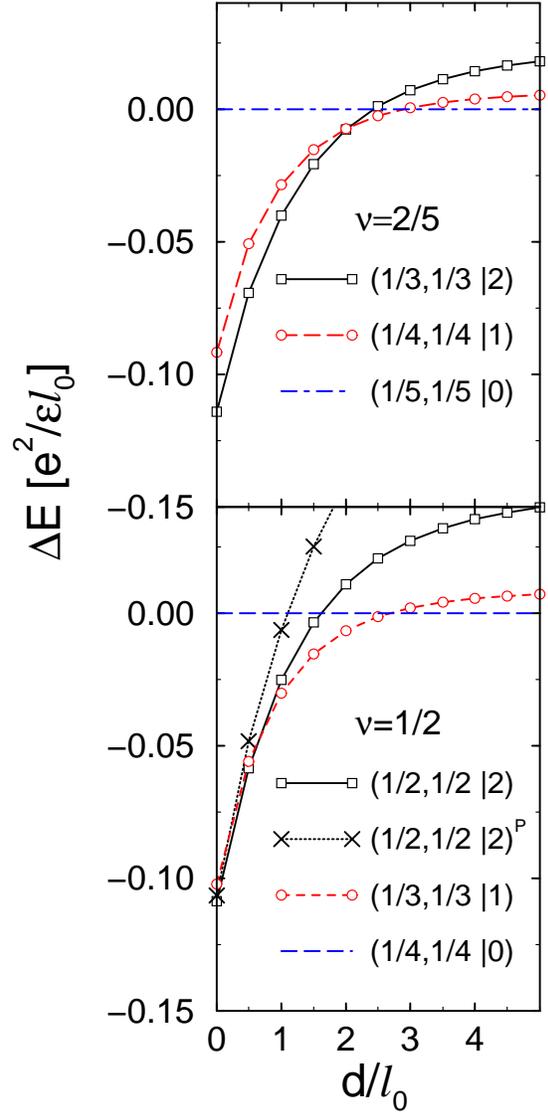,width=5in,height=7in,angle=0}}
\caption{The same as Fig.~\ref{coul_1_3_1} for fillings $\nu=2/5$ and $1/2$.}
\label{coul_2_5_1_2}
\end{figure*}

Figs.~(\ref{coul_1_3_1}) and (\ref{coul_2_5_1_2}) show the energy difference 
$\Delta E=E_{(\overline{\nu},\overline{\nu}|m)}-E_{(\nu/2,\nu/2|0)}$
of the various states at several filling factors
as a function of the layer separation, $d/l_0$, with each layer taken to be 
a strictly two-dimensional electron gas.  
It is clear that {\em all} states of the 
type $(\overline{\nu},\overline{\nu}|m)$
are in general physically relevant, provided $m\leq m_c$.
For $\nu=1/3$, we 
have $m_c=3$; for $\nu=2/5$, 1/2,
3/7, we have $m_c=2$; and for $\nu=1$, $m_c=1$. 
For large $d/l_0$, the ground state, as expected, is
$(\nu/2,\nu/2,|0)$, and as $d/l_0$ is decreased, $m$ increases in steps of unity.  
For the states studied here, the transition
occurs from an incompressible state into a compressible state and {\em vice
versa}.  This, however, is not a general feature.  For example, the bilayer states 
at $\nu=4/9$ are $(2/9,2/9|0), (2/7,2/7|1),$ and $(2/5,2/5|3)$ which are all incompressible.

At a fixed filling factor, as $d/l_0$ is varied, many transitions are predicted
to occur, with FQHE disappearing for a while and then reappearing.  This is 
reminiscent of the case of transitions
between CF states with different spin-polarizations in a single layer as a function of the
Zeeman energy.  There, however, the transition is always 
believed to be from one incompressible state to another.

\begin{figure}
\centerline{\psfig{figure=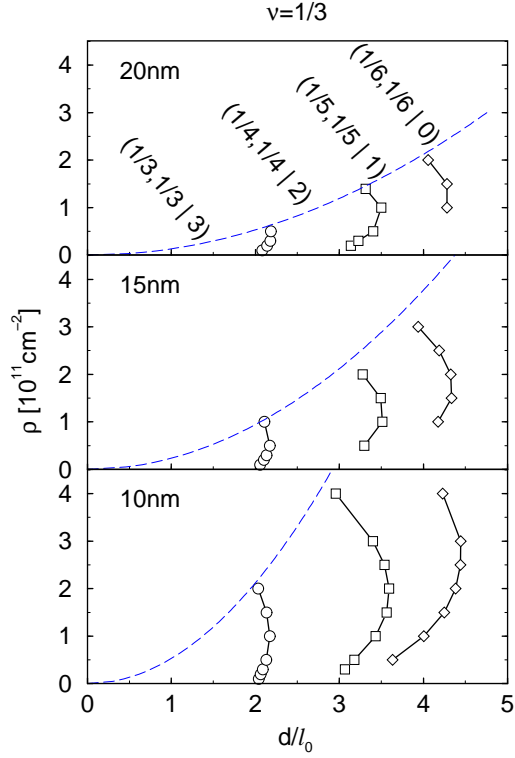,height=5in,angle=0}}
\caption{The phase diagram of bilayer CF states at $\nu=1/3$ 
in the $\rho-d$ plane, where $\rho$ is the total
electron density and $d$ is the distance between the two layers 
(measured from center to center).  The phase diagram is given for three quantum well 
widths, 10nm, 15nm, and 20nm. The dashed line shows the minimum distance between 
the layers, which is equal to the width of the individual single quantum well layer.}
\label{phase_1_3}
\end{figure}

\begin{figure}
\centerline{\psfig{figure=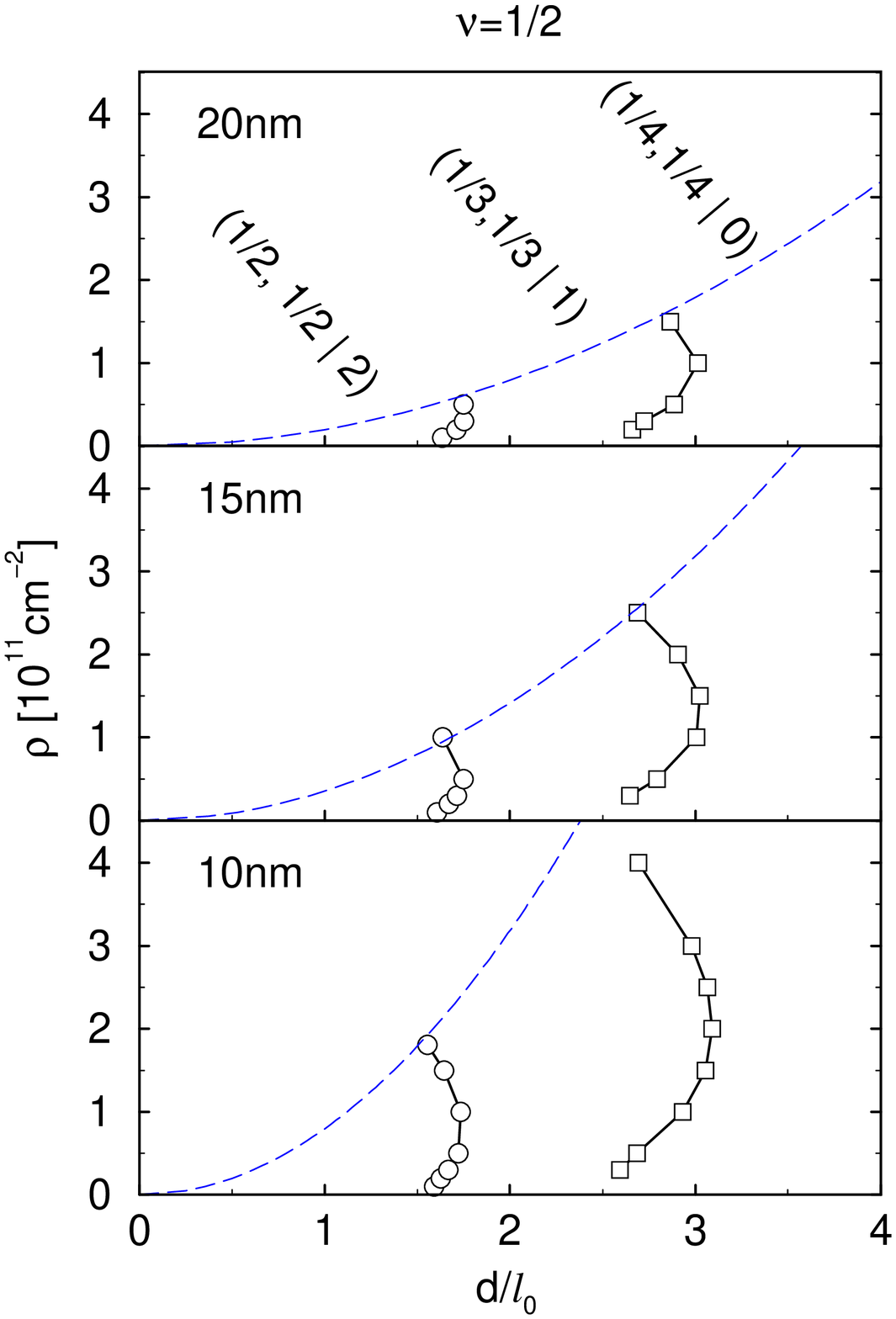,height=5in,angle=0}}
\caption{The phase diagram of bilayer CF states at $\nu=1/2$ in the $\rho-d$ plane.}
\label{phase_1_2}
\end{figure}

\begin{figure}
\centerline{\psfig{figure=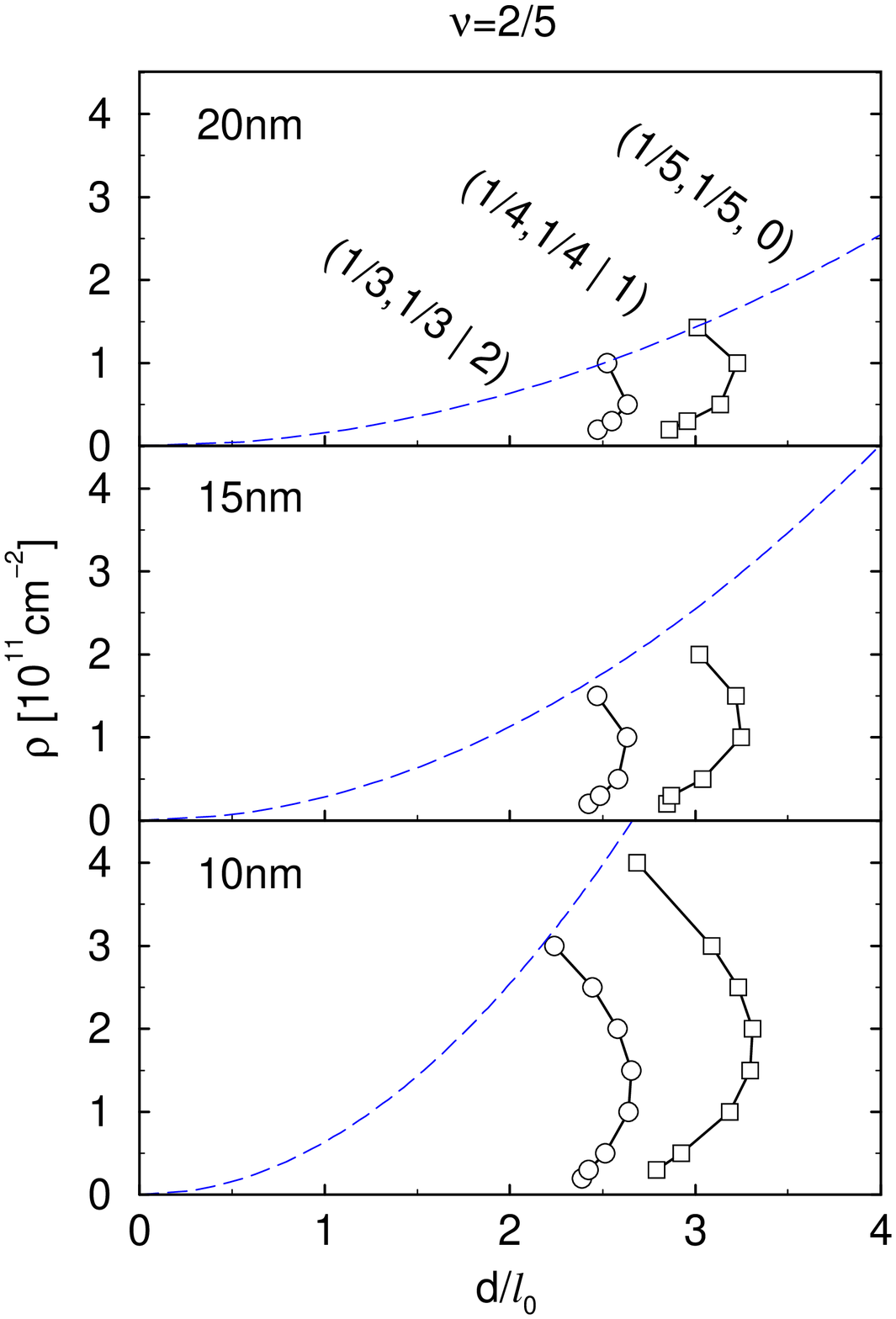,height=5in,angle=0}}
\caption{The phase diagram of bilayer CF states at $\nu=2/5$ in the $\rho-d$ plane.}
\label{phase_2_5}
\end{figure}

\begin{figure}
\centerline{\psfig{figure=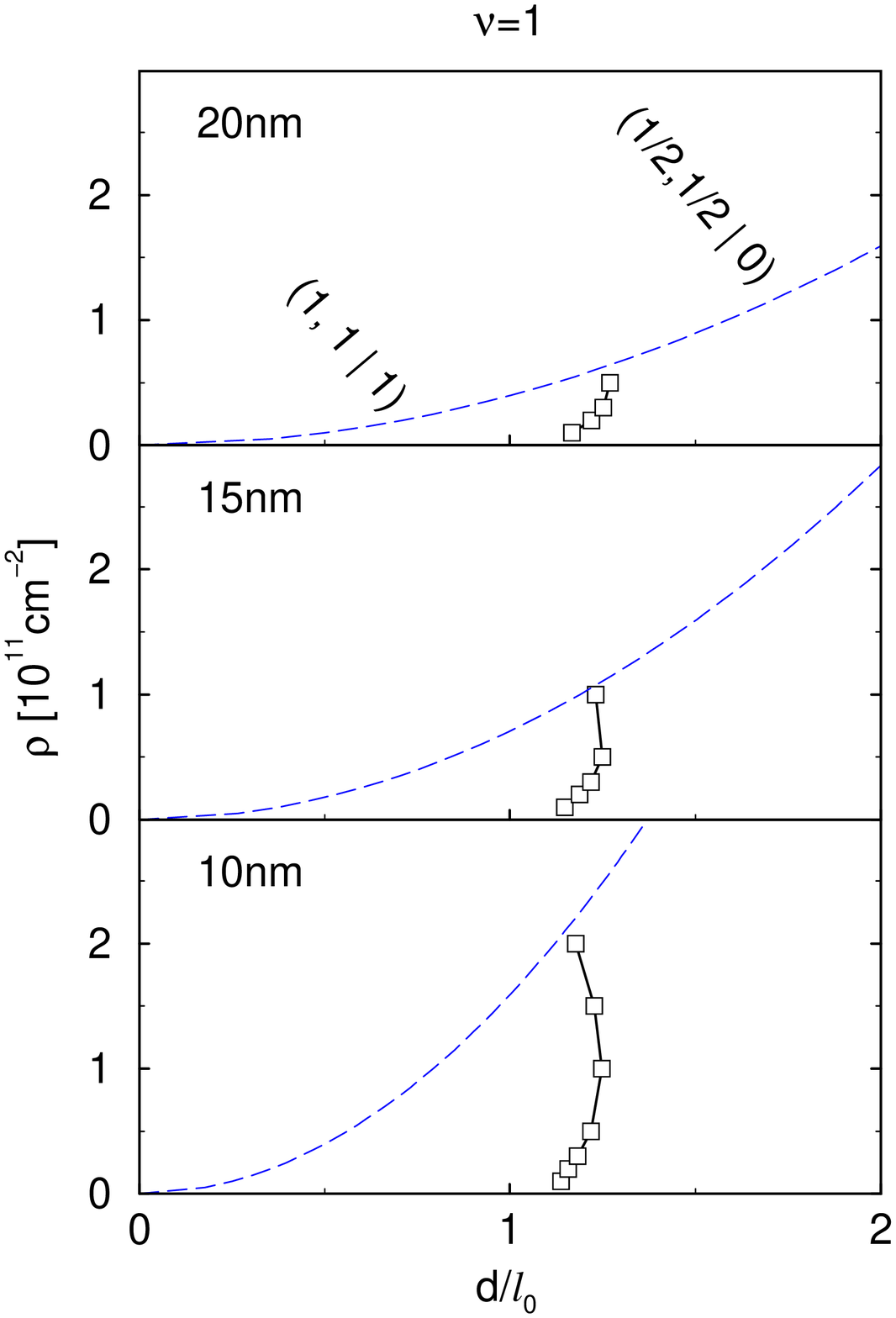,height=5in,angle=0}}
\caption{The phase diagram of bilayer CF states at $\nu=1$ in the $\rho-d$ plane.}
\label{phase_1}
\end{figure}

In order to make contact with experiment, we have considered the effects of finite width.  For
each value of the finite width, a crossover value of $d/l_0$ 
is determined as in Figs.~(\ref{coul_1_3_1}) and (\ref{coul_2_5_1_2}), 
from which the phase diagram is obtained.
The phase diagrams are shown in Figs.~(\ref{phase_1_3}),(\ref{phase_1_2}),(\ref{phase_2_5}), and (\ref{phase_1}).  The transitions occur in the 
region $d/l_0\approx 1-3$, which is an experimentally accessible
parameter range.  In recent experiments\cite{Spielman}, values as low as 1.45 for 
the ratio $d/l_0$ have been obtained by gating the sample and varying the density.
For zero thickness, the phase boundaries in the $\rho-d$ plane would be vertical.

It must be stressed that, due to its variational nature, our study cannot rule out other
phases.  For example, the bilayer quantum Wigner crystal phase has not been 
considered\cite{Manoharan}.
However, we believe that the wave functions considered in this work 
exhaust the likely uniform liquid states.

\section{Comparison with experiment and exact diagonalization studies}

Some of the FQHE states have been investigated in the past in exact diagonalization 
studies\cite{Chakraborty,Yoshioka,He,He2}.  
These studies typically deal with a total of $2N=6$ electrons, i.e.
three electrons per layer; the enlarged Hilbert space due to the additional layer 
index makes it
difficult to increase the number of particles in the exact diagonalization study 
of the bilayer problem.  As a result, it is not clear to what extent the results 
are indicative of the thermodynamic limit.  Also, these studies are not able to
effectively investigate compressible states, or higher order incompressible FQHE states.

As for the incompressible states, there is general agreement between our study
and the earlier ones, whenever a comparison is possible.  For example, the regions 
of stability of $(1/5, 1/5|0)$ and $(1/3, 1/3|2)$ in
Ref.~\onlinecite{Yoshioka} are in rough agreement with ours.  At 
$\nu=1/2$, both Refs.~\onlinecite{Yoshioka,He} find that the largest overlap between the
exact state and the Halperin $(1/3, 1/3 | 1)$ state occurs at around $d/l_0\approx 1.5$, 
which lies in the parameter range where we find the $(1/3, 1/3 | 1)$ to be stable.
At $\nu=1$, our Coulomb result $d_c/l_0\approx 1.2$ compares well with $d_c/l_0\approx 1.3$ 
obtained in Ref.~\onlinecite{Schliemann};  they find, however, a stronger dependence with the
quantum well width than we do.  (Part of the difference might arise from their 
use of a sine function for the transverse wave functions as opposed to our self-consistent
LDA.) In some cases, however, there are quantitative differences.
For example, Yoshioka, MacDonald, and Girvin \cite {Yoshioka} 
find (for a pure two-dimensional system) 
that at $\nu=1/3$, the $(1/5, 1/5|1)$ state is stable in the region $1.5 < d/l_0 < 4$, as
opposed to our study that finds the regime of stability to be $3.0<d/l_0<4.0$. 
Or, for the $(1/3, 1/3 | 1)$ state at $\nu=1/2$, the overlap does not diminish 
substantially even for $d/l_0$ up to 4 \cite{Yoshioka,He}.

The table I compares our theoretical results with experiments.
For $\nu=1/2$, the theoretical value $d/l_0\approx 3.0$ is in excellent agreement with
experiment \cite{Eisenstein}. (Note that the geometry studied
in this work is closer to the experiment of Ref.~\onlinecite{Eisenstein} than of
Ref.~\onlinecite{Suen}.) However, for $\nu=1$, the theoretical value ($d/l_0\approx 1.2$) 
is substantially smaller than the experimental values ($d/l_0\approx 1.7-2.0$).  There may be
several possible origins of this discrepancy.  The most likely cause 
is that at smaller values of
$d/l_0$, the effect of inter-layer tunneling, neglected in theory, is not negligible.
Indeed, Schliemann {\em et al.} \cite{Schliemann} find that the critical $d_c/l_0$ increases
with the tunneling gap.  Also, when the two layers are close, the inter-layer interaction may
also affect the form of the LDA form of $\xi$.

Another interesting feature of our results is that there is a range of parameters, with
$d/l_0\leq 2$, where the $\nu=n/(2n+1)$ FQHE is stable and at the same time the $\nu=1/2$ FQHE
also occurs.  It may seem that the system behaves as a single layer except that it also shows
FQHE at $\nu=1/2$.  This has been seen experimentally \cite{Shayegan}. However, 
the nature of the $\nu=n/(2n+1)$ FQHE is quite different from that
in a single layer, which ought to be verifiable from the detailed properties of this state.

Another case of interest is $\nu=2/3$.  Here, the relevant bilayer states are $(1/3,1/3|0)$, 
$(1/2,1/2|1)$, and $(1,1|2)$.  The last one needs some explanation.  Naively, it might seem
unphysical, because it appears to have stronger interlayer correlations than intralayer
correlations.  However, we note that $\psi_1$ here is not $\Phi_1$, but rather
$\prod_{j<k}(z_j-z_k)^2\Phi_{-1}$.  In other words, the 1 in $(1,1|2)$ refers to the state 
$\overline{n}/(2\overline{n}+1)$ with $\overline{n}=-1$; 
it is the $\nu^*=-1$ state of composite fermions
carrying two flux quanta.  This is the valid pseudospin-singlet  state in the $d/l_0=0$ limit
\cite{Wu}.  We have not studied $\nu=2/3$ in this work because $(1,1|2)$ involves reverse
flux attachment, for which our method for treating the lowest Landau level projection 
does not work, for technical reasons.  However, it appears that while the bilayer state at
$\nu=2/3$ is incompressible for the two extreme limits of far separated and very close layers,
a compressible state is to be expected in an intermediate range of layer separation. 
In the experimental studies of Suen {\em et al.}\cite{Shayegan} and Manoharan {\em et al.} 
\cite{Manoharan}, there is evidence for a direct transition between a two component 
incompressible 2/3 state to a single component, incompressible
2/3 state as a function of the interlayer tunneling (which
determines the gap between the symmetric and antisymmetric bands).  Our theory, on the other
hand, neglects any interlayer tunneling, and deals only with various kinds of 
two-component states.  It is not possible to draw any definitive conclusions in this 
regard from exact diagonalization studies\cite{He}.

\section{Paired Composite-fermion States}

As mentioned earlier the variational nature of our study does not exclude 
other candidate states.  To this end we consider another class of states that involve 
intra-layer pairing of composite fermions. Such a pairing is 
described by a Pfaffian variational wave function \cite{Moore},  studied by
several groups in the context of single layer FQHE at $\nu=1/2$ and $\nu=5/2$
\cite{Greiter,Park}. The Pfaffian wave function at $\overline {\nu}=1/2p$ is written as
\begin{equation}
\psi_{1/2p}^{Pf}[\{z_j\}]=\prod_{i<j}(z_i-z_j)^{2p}Pf[M] 
\end{equation}
where $Pf[M]$ is the Pfaffian of the $N\times N$ antisymmetric matrix $M$ with 
components $M_{ij}=(z_i-z_j)^{-1}$, defined as
\begin{equation}
Pf[M]\propto A[M_{1,2} M_{3,4} ... M_{N-1,N}]
\end{equation}
where $A$ is the antisymmetrization operator.  
$Pf[M]$ is a real space BCS wave function and so $\psi_{\overline{\nu}}^{Pf}$ 
can be viewed as a $p$-wave paired quantum Hall state of composite fermions.  We generalize the 
single layer paired state to the case of a bilayer system by substituting 
$\psi_{\overline{\nu}}^{Pf}$ into Eq.~\ref{Psi}:
\begin{equation}
\Psi_{(1/2p,1/2p|m)^P}^{Pf}= \prod_{r,j}(z_j-w_r)^m 
\psi_{1/2p}^{Pf}[\{z_k\}] \psi_{1/2p}^{Pf}[\{w_s\}]
\label{pf}
\end{equation}
These states will be denoted by $(1/2p,1/2p|m)^P$.
It is well known that at $\nu=1/2p$ in the single layer case the composite-fermion
Fermi sea has lower energy than the paired composite fermion state in the lowest Landau level.
\cite{Park,Greiter}  Here we check whether or not inter-layer correlations 
stabilize the intra-layer paired states and thus energetically favor the 
paired bilayer state.  

We have calculated the thermodynamic energies of the possible paired bilayer states
at total fillings $\nu=1/2, 1/3, 1,$ and $2/5$ as a function of the inter-layer
width, $d/l_0$.  The crosses in 
Figs.~(\ref{coul_1_3_1}) and (\ref{coul_2_5_1_2}) show the energy 
of the paired bilayer states measured relative to the  
uncoupled $m=0$ bilayer states of Eq.~\ref{states} as a function 
of $d/l_0$.  [We include in Figs.~(\ref{coul_1_3_1}) and (\ref{coul_2_5_1_2})
only those paired states which are competitive with 
the CF bilayer states.]  On the basis of these studies, we conclude that the 
paired CF states of Eq.~\ref{pf} are not the ground states at any layer separation 
for any of the filling factors considered.

\section{Conclusion}

We have generalized the composite fermion theory to bilayer systems in the absence of
interlayer tunneling.  We have studied a class of wave functions at primary bilayer 
fractions and 
made detailed theoretical predictions regarding the phase diagram of the bilayer states in the
density-layer separation plane.  One feature that comes out of our study is that 
the phase diagram often contains alternating stripes of compressible and incompressible phases.
Our results are generally consistent with previous theoretical and experimental studies
whenever a comparison could be made.

This work was supported in part by the National Science Foundation under grants no. 
DMR-9986806 and DGE 9987589.  We are grateful to the Numerically Intensive Computing 
Group led by V.
Agarwala, J. Holmes, and J. Nucciarone, at the Penn State University CAC for
assistance and computing time with the LION-X cluster, and thank Kwon Park and Sudhansu 
Mandal for helpful discussions.

\begin{table*}[t]
\caption{The experimental and theoretical values for the critical bilayer separation $d_c/l_0$
where the transition takes place at $\nu=1/2$ and $\nu=1$.
\label{tab:Tab1}}
\vspace{0.4cm}
\begin{tabular}{|c|c|c|c|} 
$\nu$   &  $d_c/l_0$ (experiment) & $d_c/l_0$ (present theory) & Reference \\ \hline
1  &    2.0 &      1.2 &       \onlinecite{Murphy}\\ \hline
1 &     1.7 &      1.2 &       \onlinecite{Hyndman} \\ \hline
1  &    1.8  &     1.2   &    \onlinecite{Hamilton} \\ \hline
1/2 &   3.0  &     2.9 &     \onlinecite{Eisenstein}\\ 
\end{tabular}
\end{table*}

\begin{references}
\bibitem{footnote2} Note that negative values
of $n$ are also allowed, with $\Phi_{-n}=[\Phi_n]^*$.  These correspond to reverse flux
attachment, in which the effective magnetic field for composite fermions 
points in a direction opposite to the external magnetic field.
\end{references}
\end{document}